\documentclass[a4paper,preprintnumbers,floatfix,
superscriptaddress,prl,onecolumn,showpacs,
notitlepage,longbibliography]{revtex4-1}

\usepackage[utf8x]{inputenc}
\usepackage{graphicx}
\usepackage{dcolumn}
\usepackage{bm}
\usepackage{braket}
\usepackage{amsmath}
\usepackage{amstext}
\usepackage{amssymb}

\begin{document}


\title{Greenberger-Horne-Zeilinger state generation with linear optical elements}



\author{Bert\'ulio de Lima Bernardo}
 \affiliation{Departamento de F\'isica, Universidade Federal da Para\'iba, 58051-900 Jo\~ao Pessoa, PB, Brazil}
\affiliation{%
Departamento de Física, Universidade Federal de Campina Grande, 58429-900, Campina Grande, Paraíba, Brazil}
\author{M. Lencs\'es}%
\author{S. Brito}
\affiliation{%
 International Institute of Physics, Federal University of Rio Grande do Norte, 59070-405 Natal, Brazil
}%

\author{Askery Canabarro}
\email{askery.canabarro@arapiraca.ufal.br}
\affiliation{%
 International Institute of Physics, Federal University of Rio Grande do Norte, 59070-405 Natal, Brazil
}%
\affiliation{%
 Grupo de F\'isica da Mat\'eria Condensada, N\'ucleo de Ci\^encias Exatas - NCEx, Campus Arapiraca, Universidade Federal de Alagoas, 57309-005 Arapiraca-AL, Brazil
}%




\date{\today}

\begin{abstract}
We propose a scheme to probabilistically generate Greenberger-Horne-Zeilinger (GHZ) states encoded on the path degree of freedom of three photons. These photons are totally independent from each other, having no direct interaction during the whole evolution of the protocol, which remarkably requires only linear optical devices to work, and two extra ancillary photons to mediate the correlation. The efficacy of the method, which has potential application in distribited quantum computation and multiparty quantum communication, is analyzed in comparison with similar proposals reported in the recent literature. We also discuss the main error sources that limit the efficiency of the protocol in a real experiment and some interesting aspects about the mediator photons in connection with the concept of spatial nonlocality. 
%
\end{abstract}


\maketitle


\section{\label{sec:1}Introduction}

As famously quoted by Schr\"{o}dinger, entanglement is not one but rather the characteristic trait of quantum mechanics \cite{schro}, a cornerstone in our modern understanding of quantum theory giving rise to fundamental and practical applications. Entanglement not only is a central concept in quantum information science \cite{horo} but more recently has also attracted increasing attention in the most various fields \cite{EntanglementQFT,EntanglementBio,EntanglementBlack,EntanglementCondensed}. From the more applied perspective, entanglement has been recognized as the key resource in many quantum information protocols, ranging --just to mention the most paradigmatic examples-- from quantum teleportation \cite{bennett} to quantum dense coding \cite{bennett2}, quantum cryptography \cite{ekert}, and quantum computation \cite{nielsen}.

Given its importance, not surprisingly, the experimental generation of entangled states has been extensively studied both from the theoretical and experimental sides in a number of different physical systems, such as superconducting circuits \cite{carlo}, trapped ions \cite{blatt}, nuclear magnetic resonance systems \cite{nelson}, and photons \cite{lu}. In particular, the emblematic bipartite entangled Bell state now is achieved
routinely with fidelities close to unity \cite{BellHigh}. The multipartite case, however, is still very challenging experimentally. Not only, the multiple parties give rise to infinitely many inequivalent classes of entanglement \cite{dur,EntanglementInfinity}, its experimental generation is very sensible to noise, typically leading to fidelities and probabilities of success that scale badly as we increase the system's size \cite{EntanglementGHZ,Entanglementdecay,GHZgen}. In spite of that, over the years the experimental challenges have steadily been surpassed, with the controlled generation of multipartite states reaching impressive 20 entangled particles in a fully controllable manner \cite{Entanglement20}.

The paradigmatic example of a multipartite entangled state is the so-called Greenberger-Horne-Zeilinger (GHZ) state \cite{green}, that in the particular case of 3 parties takes the form
\begin{equation}
\label{1}
\ket{GHZ} = \frac{1}{\sqrt{2}}(\ket{000}+\ket{111}),
\end{equation}    
where $\ket{0}$ and $\ket{1}$ are orthogonal states. Its importance stems from its wide variety of applications, in protocols such as controlled dense coding \cite{hao}, quantum metrology \cite{Metro,Metronoise}, synchronization of clocks \cite{Clock} and measurement based quantum computation \cite{MBQC}. While GHZ states have been experimentally generated in a few different physical setups, in applications involving large distances such as quantum secret sharing \cite{hil} or Bell inequalities violation \cite{Bellrev}, photonic platforms are particularly advantageous given the relative ease of transmission and robustness of photons. Typically, in photonic implementations, the quantum information is encoded in the polarization degree of freedom \cite{bou}, mostly because of its simple local manipulation requiring only linear optical elements. Polarization, however, also has its drawbacks. First, it becomes delicate when optical fibers for long-distance applications are required because of the unavoidable polarization perturbations upon propagation \cite{bert,preskill}. Second, given the low efficiency of the photon detectors upon which polarization measurements are performed, often one has to rely on post-selection of data, a critical issue, for instance, in cryptographic applications \cite{ekert}. Thus, it becomes clear the relevance of considering GHZ states encoded in different degrees of freedom, precisely what we pursue here.


In this paper, we propose a new method to probabilistically generate path-encoded GHZ states with photons. A remarkable feature of it, contrary to previous proposals \cite{berg}, is that it only requires linear optical elements. Similar to other protocols involving GHZ states \cite{hao,su,ri}, the revelation of the generated state only occurs by postselection. Another interesting feature of our protocol is that the photons to be entangled do not interact directly with each other during the state preparation, in a process akin to entanglement swapping \cite{Eswap}. That is, they remain spatially separated for all times. Moreover, the photons need no previous entanglement, thus explaining the total exemption of nonlinear optical elements, as for example those required in a parametric down-conversion process. The paper is organized as follows. In Sec. II we expose the main idea of the protocol and the probabilities to generate GHZ and GHZ-class states, which we shall call desired states. In Sec. III, we study the entanglement properties of the desired states. In Sec. IV, we investigate the main sources of errors which are present in our method and discuss how they affect the generation rate of GHZ states. We conclude with a summary of the results in Sec. V.


\section{\label{sec:2}Generation of GHZ states}

In this section we develop the protocol to generate tripartite GHZ states, without considering any source of errors, which will be considered in Section IV. The protocol is conceptually similar to the ideas developed in \cite{bert1,bert2} for the generation of bipartite Bell states. It also bears a conceptual link with what happens in an entanglement swapping experiment \cite{Eswap}. More specifically, we propose to entangle three distant particles -- which have no common past and never interact directly -- by projective measurements on two other particles. As depicted in Fig. 1, the entire setup consists of five single photon-sources, five Mach-Zehnder(MZ)-like apparatuses (composed of a total of eleven beam splitters (BS) and two mirrors (M)) and 4 bulk photon detectors.

The first 5 beam splitters (BS$_{i}$, with $i=1,\dots,5$), convert the state of inputs photons into a superposition of the ten different spatial modes denoted by the state $\ket{A}_{(1,2,3,4,5)}$ and $\ket{B}_{(1,2,3,4,5)}$. The arms length of each of the modes must be such that the arrival time of each of the photons coincide at the secondary layer of four beam splitters and 2 mirrors (BS$_{6}$, BS$_{7}$, BS$_{8}$, BS$_{9}$, M$_{1}$ and M$_{2}$). In this case, whenever two of the photons arrive at the opposite ports of this second layer of beam splitters (BS$_{6}$, BS$_{7}$, BS$_{8}$, BS$_{9}$), the Hong-Ou-Mandel (HOM) effect occurs in such a way that the photons involved necessarily exit the beam splitter through the same output port \cite{hong}. As seen in Fig. 1, after this second layer of beam splitters, no other transformation is performed on spatial modes $\ket{A}_{(1,3,5)}$ and $\ket{B}_{(1,3,5)}$. However, paths $\ket{A}_{(2,4)}$ and $\ket{B}_{(2,4)}$ impinge in a final layer of two beam splitters(BS$_{10}$ and BS$_{11}$) before detection.
\begin{figure}[ht]
\begin{center}
\includegraphics{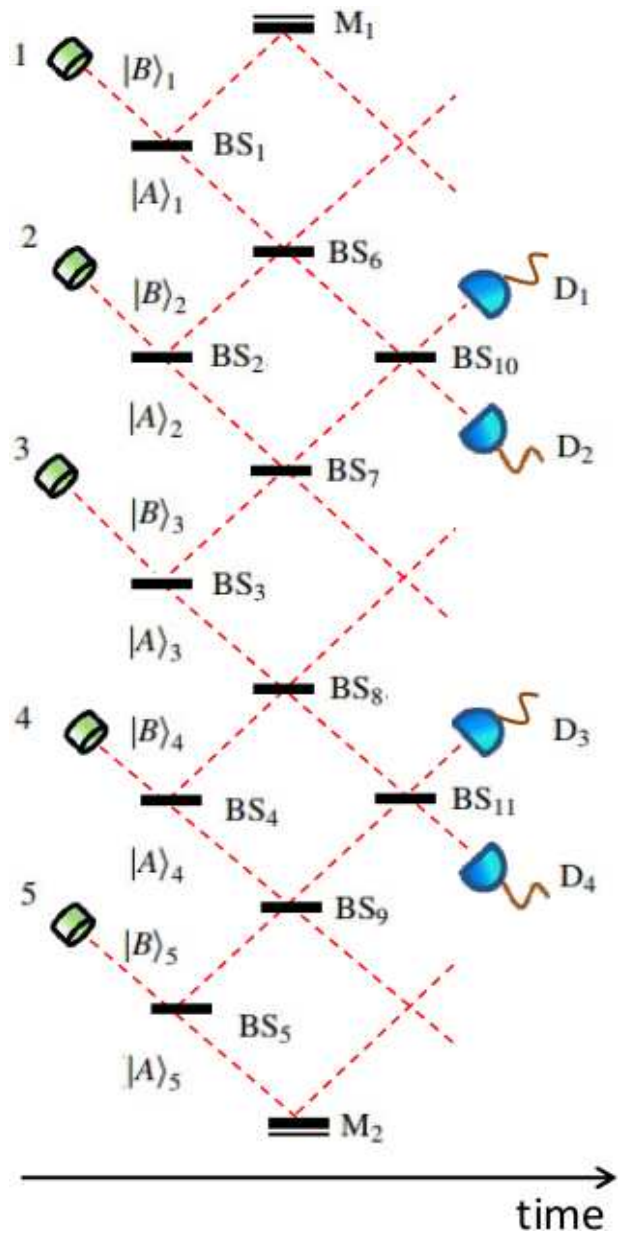}
\end{center}
\label{F2}
\caption{Sketch of the GHZ state generator setup which uses only linear optical elements. Five independent photons 1, 2, 3, 4 and 5 enter five Mach-Zehnder-like circuits, which share four central beam splitters, BS$_{6}$, BS$_{7}$, BS$_{8}$ and BS$_{9}$. When we consider only the cases in which no transmission takes place at these central beam splitters, depending on which combination of clicks in the four possible detectors, photons 1, 3 and 5 are transformed into four possible genuine tripartite entangled states at the end of the process. One of the possibilities is the desired GHZ state, which takes place under the coincident detection of photons 2 and 4 at D$_{1}$ and D$_{4}$, respectively. The time arrow specifies the exact sequence of steps taken in the method.}
\end{figure}

After the first layer of beam-splitters the state is given by
\begin{eqnarray}
\label{2}
\ket{I} &=& \frac{1}{4 \sqrt{2}}(\ket{A}_{1}+i\ket{B}_{1})\otimes(\ket{A}_{2}+i\ket{B}_{2})\\ \nonumber
&\otimes&(\ket{A}_{3}+i\ket{B}_{3})\otimes(\ket{A}_{4}+i\ket{B}_{4})\otimes(\ket{A}_{5}+i\ket{B}_{5}),
\end{eqnarray}   
where we have considered that the photons acquire no phase by transmission through a beam splitter, however, a phase $i$ is obtained upon reflections. An essential point for the realization of the protocol is that we only consider (post-select) the cases in which each photon remains in its respective circuit after each run of the experiment. In other terms, no photon arrives at the opposite ports of the second layer of beam-splitters and thus we can identify each initial photon $i$ with its input modes $\ket{A}_i$ and $\ket{B}_i$. That is, we neglect all runs in which one photon invade the adjacent circuit, rendering three, two or none of the photons at the output ports of a given circuit. This corresponds to the terms $\ket{A}_{1}\ket{B}_{2}$, $\ket{A}_{2}\ket{B}_{3}$, $\ket{A}_{3}\ket{B}_{4}$ and $\ket{A}_{4}\ket{B}_{5}$ in Eq.~\eqref{2}, which evolve as $\ket{A}_{1}\ket{B}_{2} \rightarrow \frac{i}{\sqrt{2}}(\ket{2A_{1}}+\ket{2B_{2}})$, $\ket{A}_{2}\ket{B}_{3} \rightarrow \frac{i}{\sqrt{2}}(\ket{2A_{2}}+\ket{2B_{3}})$,  $\ket{A}_{3}\ket{B}_{4} \rightarrow \frac{i}{\sqrt{2}}(\ket{2A_{3}}+\ket{2B_{4}})$ and $\ket{A}_{4}\ket{B}_{5} \rightarrow \frac{i}{\sqrt{2}}(\ket{2A_{4}}+\ket{2B_{5}})$, as a result of the two-photon interference effect at the central beam splitters. Upon post-selection we can exclude such  events. Thus, the following terms remain from the state $|I\rangle$ of Eq. \ref{1}:

\begin{eqnarray}
\label{3}
\ket{I} \rightarrow & & \frac{|A\rangle _1|A\rangle _2|A\rangle _3|A\rangle _4|A\rangle _5}{4 \sqrt{2}}+\frac{i |B\rangle _1|A\rangle _2|A\rangle _3|A\rangle _4|A\rangle _5}{4 \sqrt{2}} \nonumber \\
& &-\frac{|B\rangle _1|B\rangle _2|A\rangle _3|A\rangle _4|A\rangle _5}{4 \sqrt{2}}-\frac{i |B\rangle _1|B\rangle _2|B\rangle _3|A\rangle _4|A\rangle _5}{4 \sqrt{2}} \nonumber \\& &+\frac{|B\rangle _1|B\rangle _2|B\rangle _3|B\rangle _4|A\rangle _5}{4 \sqrt{2}}+\frac{i |B\rangle _1|B\rangle _2|B\rangle _3|B\rangle _4|B\rangle _5}{4 \sqrt{2}}. \nonumber \\
\end{eqnarray} 
    
Since we are neglecting the cases in which photons invade the neighbor circuits, we have that reflections must occur in all secondary devices, rendering a phase factor $i$ in each single photon mode. Also, for us to disconsider transmissions at the central beam splitters, each incident mode must acquire a reduced amplitude factor of $1/\sqrt{2}$ due to the 50\% of reflection probability. Overall, the following rules are applicable
\\

$|A\rangle_j \rightarrow \frac{i}{\sqrt{2}}|A\rangle_j
$, for $j = 1,2,3,4$. For $ j= 5$ we have a mirror, so $|A\rangle_5 \rightarrow i|A\rangle_5, $
\\

$|B\rangle_j \rightarrow \frac{i}{\sqrt{2}}|B\rangle_j $, for $j = 2,3,4,5$. For $j = 1$ we have a mirror, so $|B\rangle_1 \rightarrow i|B\rangle_1 $.
\\

If the above rules are applied, after {\it normalization}, we obtain the following state from Eq.~\eqref{2}:

\begin{eqnarray}
\label{4}
\ket{\psi} &=& \frac{i |A\rangle _1|A\rangle _2|A\rangle _3|A\rangle _4|A\rangle _5}{\sqrt{10}} -\frac{|B\rangle _1|A\rangle _2|A\rangle _3|A\rangle _4|A\rangle _5}{\sqrt{5}} \nonumber \\
&-&\frac{i |B\rangle _1|B\rangle _2|A\rangle _3|A\rangle _4|A\rangle _5}{\sqrt{5}}+\frac{|B\rangle _1|B\rangle _2|B\rangle _3|A\rangle _4|A\rangle _5}{\sqrt{5}} \nonumber \\
&+&\frac{i |B\rangle _1|B\rangle _2|B\rangle _3|B\rangle _4|A\rangle _5}{\sqrt{5}}-\frac{|B\rangle _1|B\rangle _2|B\rangle _3|B\rangle _4|B\rangle _5}{\sqrt{10}}. \nonumber \\
\end{eqnarray}
This is the quantum state of the five photons after the secondary devices, given the mentioned postselection. The success probability in obtaining this state from the input state $\ket{I}$ of Eq.~\eqref{1} is given by the sum of the square of the absolute values of the coefficients of the state of Eq.~\eqref{3}, {\it prior} to normalization, which yields $P \approx 0.183211$.

It is worthwhile to mention that the present protocol with the setup of Fig. 1 and the requirement that the photons remain in the original circuit during the entire process is different from the case in which the central beam splitters, BS$_{6}$, BS$_{7}$, BS$_{8}$, BS$_{9}$, are replaced by mirrors. Indeed, if we considered the latter case, the transformation caused by the secondary devices would simply be  $\ket{I} \rightarrow i \ket{I}$, where the phase factor $i$ is due to the ten possible reflections at the secondary devices, with no amplitude reduction due to the impossibility of transmissions. This transformation is obviously different from those used to obtain Eq.~\eqref{4}. 

Now, let us assume that the path lengths of MZ$_{2}$ are such that if the state of photon 2 is $(i\ket{A}_{2}+\ket{B}_{2})/ \sqrt{2}$ before arriving at BS$_{10}$ it will certainly be detected at D$_{1}$ and, conversely, if the state is $(i\ket{A}_{2}-\ket{B}_{2})/ \sqrt{2}$ before arriving at BS$_{10}$ it will certainly be detected at D$_{2}$. By the same reasoning, we will consider the configuration of MZ$_{4}$ such that if the state of photon 4 is $(i\ket{A}_{4}-\ket{B}_{4})/ \sqrt{2}$ before arriving at BS$_{11}$ it will be detected at D$_{4}$, whereas if under the same circumstance the state is $(i\ket{A}_{4}+\ket{B}_{4})/ \sqrt{2}$ photon 4 will be detected at D$_{3}$.

In this regard, we define the states
\\
\begin{eqnarray}
\ket{D}_{1} &=& (i\ket{A}_{2}+\ket{B}_{2})/ \sqrt{2}, \nonumber \\
\ket{D}_{2} &=& (i\ket{A}_{2}-\ket{B}_{2})/ \sqrt{2}, \nonumber \\
\ket{D}_{3} &=& (i\ket{A}_{4}+\ket{B}_{4})/ \sqrt{2}, \nonumber \\
\ket{D}_{4} &=& (i\ket{A}_{4}-\ket{B}_{4})/ \sqrt{2},
\end{eqnarray}
that, after the secondary devices, correspond to the states in which photon 2 will be certainly detected at D$_{1}$ and D$_{2}$, respectively, and photon 4 at D$_{3}$ and D$_{4}$, respectively. In this form, we can define the projectors   
\\
\begin{eqnarray}
\Pi_{1} &=& \ket{D}_{1} \bra{D}_{1}, \nonumber \\
\Pi_{2} &=& \ket{D}_{2} \bra{D}_{2}, \nonumber \\
\Pi_{3} &=& \ket{D}_{3} \bra{D}_{3}, \nonumber \\
\Pi_{4} &=& \ket{D}_{4} \bra{D}_{4},
\end{eqnarray}
which will assist us in calculating the state of photons 1, 3 and 5, upon detection of photons 2 and 4 at any combination of detectors. In fact, if we are interested in the state of photons 1, 3 and 5, when photons 2 and 4 are detected at D$_{1}$ and D$_{3}$, D$_{1}$ and D$_{4}$, D$_{2}$ and D$_{3}$, and D$_{2}$ and D$_{4}$, respectively, after normalization, it can be found to be   

\begin{equation}
\label{7}
\ket{\psi}_{1} = \Pi_{1} \Pi_{3} \ket{\psi}=-\frac{i |A\rangle _1 |A\rangle _3 |A\rangle _5}{\sqrt{2}}-\frac{|B\rangle _1 |B\rangle _3 |B\rangle _5}{\sqrt{2}}
\end{equation}

\begin{equation}
\label{8}
\begin{split}
\ket{\psi}_{2} = \Pi_{1} \Pi_{4} \ket{\psi} = -\frac{i |A\rangle _1 |A\rangle _3 |A\rangle _5}{\sqrt{10}}-\frac{2 i |B\rangle _1 |B\rangle _3 |A\rangle _5}{\sqrt{5}}\\+\frac{|B\rangle _1 |B\rangle _3 |B\rangle _5}{\sqrt{10}}
\end{split}
\end{equation}


\begin{equation}
\label{9}
\begin{split}
\ket{\psi}_{3} = \Pi_{2} \Pi_{3} \ket{\psi} =-\frac{i |A\rangle _1 |A\rangle _3 |A\rangle _5}{\sqrt{10}}+ \frac{2 |B\rangle _1 |A\rangle _3 |A\rangle _5} {\sqrt{5}}\\+\frac{|B\rangle _1 |B\rangle _3 |B\rangle _5}{\sqrt{10}},
\end{split}
\end{equation}

\begin{equation}
\label{10}
\begin{split}
\ket{\psi}_{4} = \Pi_{2} \Pi_{4} \ket{\psi} =-\frac{i |A\rangle _1 |A\rangle _3 |A\rangle _5}{3 \sqrt{2}}+ \frac{2}{3} |B\rangle _1|A\rangle _3 |A\rangle _5 \\+\frac{2}{3} i |B\rangle _1 |B\rangle _3 |A\rangle _5-\frac{|B\rangle _1 |B\rangle _3 |B\rangle _5}{3 \sqrt{2}},
\end{split}
\end{equation}
whose probabilities of detection are given respectively by

\begin{equation}
P_{1} = \braket{\psi|\Pi_{1} \Pi_{3}|\psi} = 0.05,
\end{equation}

\begin{equation}
P_{2} = \braket{\psi|\Pi_{1} \Pi_{4}|\psi} = 0.25,
\end{equation}

\begin{equation}
P_{3} = \braket{\psi|\Pi_{2} \Pi_{3}|\psi} = 0.25,
\end{equation}

\begin{equation}
P_{4} = \braket{\psi|\Pi_{2} \Pi_{4}|\psi} = 0.45.
\end{equation}
A more detailed analysis of the properties of the states of Eqs.~\eqref{7} to~\eqref{10} will be provided in the next section.

\section{\label{sec:3}Entanglement properties of the possible states}


At this stage, it is important to analyze the results obtained so far. We showed that, depending on the combination in which photons 2 and 4 are detected, photons 1, 3 and 5 are launched into one of the four possible quantum states of Eqs.~\eqref{7} to~\eqref{10}, namely $\ket{\psi}_{l}$, with $l=1,2,3,4$. As a matter of fact, it is useful to study the properties of these tripartite qubit states. In doing so, we shall adopt the widely used classification proposed by D\"{u}r, {\it et. al} in Ref. \cite{dur}. That is, we will classify these states according to the equivalence classes to which they belong. These equivalence classes contain states that can be converted into each other by means of stochastic local
operations and classical communication (SLOCC). Following that analysis, we will first calculate the reduced density matrices of the states $\ket{\psi}_{l}$, say $\rho^{(l)}_{1} = Tr_{35} (\ket{\psi}_{l}\bra{\psi}_{l})$, $\rho^{(l)}_{3} = Tr_{15} (\ket{\psi}_{l}\bra{\psi}_{l})$ and $\rho^{(l)}_{5} = Tr_{13} (\ket{\psi}_{l}\bra{\psi}_{l})$, where $\rho^{(l)}_{j}$ represents the reduced density matrix of photon $j$ with respect to the tripartite state $\ket{\psi}_{l}$, and $Tr_{nm}$ the trace over the states of photons $n$ and $m$. By using these reduced density matrices, we can proceed to compute the local entropies $S^{(l)}_{j}=-Tr[\rho^{(l)}_{j} \log_{2} \rho^{(l)}_{j} ]$.  

As shown in \cite{dur}, if a given local entropy $S^{(l)}_{j}$ is null, it signifies that, in the state $\ket{\psi}_{l}$, photon $j$ is not entangled with photons $n$ and $m$. On the contrary, $S^{(l)}_{j}>0$ means that there is some amount of entanglement between $j$ and the other two photons. However, it is also known that for the case in which the state $\ket{\psi}_{l}$ is a genuine tripartite entanglement, say $S^{(l)}_{j}>0$ for $j=1,3,5$, there are still two inequivalent classes of entanglement whose constituent states cannot be obtained from each other by SLOCC. These two classes are represented by the GHZ and W states. The physical difference between these two classes is that the states pertaining to the W class retain maximally bipartite entanglement if any one of the three qubits is lost, whereas for states of the GHZ class it is impossible to maintain bipartite entanglement in an equivalent condition.

For us to distinguish states between these two genuine tripartite classes, it is necessary to calculate the 3-tangle (residual tangle), according to the recipe provided in \cite{elt}. If the 3-tangle is zero, we have that the state belongs to the W class; whereas if it is positive, it means that the state is in the GHZ-class. We performed an analysis of the tripartite entanglement for the states of Eqs.~\eqref{7} to~\eqref{10} following this classification. The results are summarized in Table 1.

\begin{center}
\begin{table}[h]
\begin{tabular}{l*{4}{c}r}
State             & $S_1$ & $S_3$ & $S_5$ & $\tau	$ & Class \\
\hline
$\ket{\psi}_1$		  & 1.000 & 1.000 & 1.000 & 1.000 & GHZ \\
$\ket{\psi}_2$        & 0.469 & 0.469 & 0.081 & 0.040 & GHZ \\
$\ket{\psi}_3$        & 0.081 & 0.469 & 0.469 & 0.040 & GHZ \\
$\ket{\psi}_4$		  & 0.187 & 0.310 & 0.187 & 0.012 & GHZ \\
\hline
\end{tabular}
\caption{Values of the local entropies, S$_{1}$, S$_{3}$ S$_{5}$, and the 3-tangle, $\tau$, of the four possible final states of photons 1, 3 and 5. The data indicate that all states belong to the GHZ class.}
\end{table}
\end{center}

As it can be seen, all four possible states obtained with the protocol of Fig. 1 are genuine tripartite entangled states pertaining to the GHZ class. However, the state $\ket{\psi}_{1}$ is the original maximally entangled GHZ state \cite{green}, after the application of a local unitary operation,

\begin{eqnarray}
\ket{GHZ} = \frac{1}{\sqrt{2}}(\ket{A}_{1}\ket{A}_{3}\ket{A}_{5}+\ket{B}_{1}\ket{B}_{3}\ket{B}_{5}) \nonumber  \\ 
=  \mathbb{I} \otimes \mathbb{I} \otimes R(-\pi/2) \ket{\psi}_{1},
\end{eqnarray}
which corresponds to two identities for photons 1 and 3, and a phase shift gate of $-\pi/2$ for photon 5.  The symbol $\doteq$ stands for equal up to an overall negative sign, which has no physical significance. With respect to the other three states, $\ket{\psi}_{2}$, $\ket{\psi}_{3}$ and $\ket{\psi}_{4}$, the fact that they are contained in the GHZ class means that they can be converted by means of SLOCC into a GHZ state. 

In  general terms, with the states obtained in Eqs.~\eqref{7} to~\eqref{10} we have that the execution of the present protocol successfully generates genuine tripartite entangled states for photons 1, 3 and 5 for all possible combinations of detection of photons 2 and 4, being also possible to generate a maximally entangled GHZ state. Furthermore, it is remarkable that such states can be created for three photons originated from independent sources. In fact, photons 1, 3 and 5 have no previous amount of entanglement, no direct interaction with each other, and even so always end up entangled. The quantum correlation established among these separated photons had necessarily its origin due to the presence of photons 2 and 4 which mediate the entanglement, in a similar fashion to the bipartite mediation processes shown in Refs. \cite{bert1,bert2}. For this reason we name the ancillary photons 2 and 4 mediators. To best of our knowledge, there is only one protocol in the literature which aims the creation of a GHZ state among three particles which never interacted, the so-called multiparticle entanglement swapping \cite{bose}, which was already realized experimentally \cite{Ylu}. However, in that scheme the GHZ state is encoded in the polarization state of photons, and the usage of nonlinear optical elements is necessary because of the prior distribution of singlet states among the distant users in the communication network.     

Particularly, one of the most intriguing aspects of the protocol is the fact that photons 1 and 5 always end up entangled, once these two particles not only are spatially separated during the whole evolution of the system, but also because they share no common mediator. Indeed, we can intuitively say that photon 2 mediated the entanglement between photons 1 and 3, and that photon 4 mediated the entanglement between photons 3 and 5. Nevertheless, the final entanglement between photons 1 and 5 is undoubtedly unexpected. Given such spatially nonlocal effect of the mediators, we believe that it is also possible to extend this mediation of entanglement to cases of higher dimensional subsystems and to systems with more than three parties.

\section{\label{sec:4}Error Evaluation of the Protocol}

In this section we investigate the influence of the two main sources of errors in the present protocol. The most important source of error that has to be mentioned is that, having two of the four detectors clicked according to our postselection rules, one still cannot be sure that the outgoing state is of the desired type. For example, the case in which only transmissions take place at all beam splitters in Fig. 1. In fact, at the end of this process, photons 1 and 3 would be respectively detected at D$_{2}$ and D$_{4}$, photon 2 would exit through the lower port of circuit 3, and photons 4 and 5 through the opposite ports of circuit 5. Thus, in this situation the desired postselection of single photons at D$_{2}$ and D$_{4}$ is misleading because we would obtain the state $\ket{B}_{3}\ket{A}_{5}\ket{B}_{5}$ instead of the desired state of Eq. (10). Other few cases of misleading postselection can also be verified. In general, it means that the protocol does not produce GHZ states on demand, a fact that takes place in a number of protocols involving GHZ states \cite{su,ri,berg}. Of course, this problem can be completely solved by verifying how many photons exit the output ports of circuits 1, 3 and 5, by means of a further postselection. If only one photon emerges from each of these circuits, it means that the protocol was successful.

Another source of error, now related to the refinement of the experimental apparatus, is the intrinsic probabilistic aspect of the success of the protocol due to the assumption of proper HOM effect realizations that, in practice, has some limitations. In this respect, it is essential that the arrivals of the photons to the second layer of beam splitters occur in such a way that their wave packets are  superposed \cite{hong}. A failure in the realizations of the HOM effect at this stage could give rise to misleading postselections, which result from independent single photon reflection or transmissions at the beam splitters. Such outcomes do not correspond to the mediation of entanglement, and hence the generation of GHZ states. For example, photons 1 and 2 can exchange their channels after reaching BS$_6$. Since we are unable to distinguish them in the the post-selection process, we would accept outcomes which are not the desired states. The influence of such misleading cases requires some analysis.

As explained in the Appendix one can treat both sources of errors in a unified way. One can formally distinguish between the five photons and label them correspondingly. Then, one can consider all possible histories with respect to the different paths after each beam splitter, keeping track of the reflections, transmissions and, therefore, possible failures of the HOM effect. At the end one can ``postselect'' the outcomes, keeping only the terms with single detector clicks, and also ``second postselect'', i.e., keeping the states which have only one particle in each port. In this way, one can obtain the most general outcome $|\Psi_{gen}(\delta)\rangle$, the postselected states $|\Psi_{p.s.,D_iD_j}(\delta)\rangle$, and the single-particle outcomes $|\Psi_{sing.p.,D_iD_j}(\delta)\rangle$ as a function of the failure probability $\delta$ of the HOM effect at the second layer of beam splitters.
In this case, $\delta=0$ means that all HOM processes succeed, and $|\Psi_{sing.p.,D_iD_j}(\delta=0)\rangle=|GHZ_{ij}\rangle$. With these general outcomes, one can calculate the overlaps between different states, which we demonstrate in Figs. 2 and 3. In Fig. 2 one can see the overlap of GHZ states with the most general outcome of the protocol as a function of $\delta$. These are the probabilities of generating GHZ states in the protocol. Fig. 3. shows the overlap between the postselected states and the GHZ state outcomes as a function of $\delta$. These results can be interpreted as the fidelities of the postselected states with respect to the desired states.

\begin{figure}[ht]
\begin{center}
\includegraphics[height=2in]{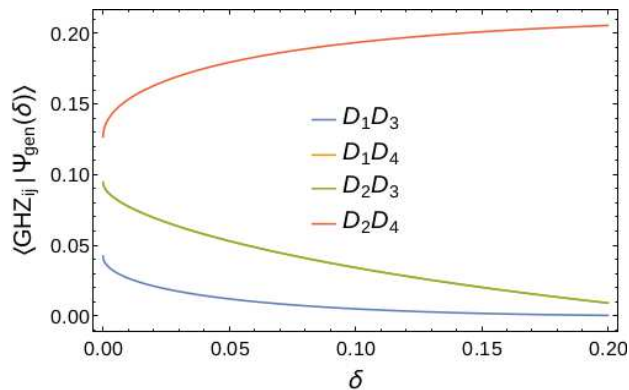}
\end{center}
\label{fig:olap_gen_ghz}
\caption{Overlaps of the different GHZ state outcomes with the most general outcome, i.e., the fidelity in generating a GHZ state as a function of the failure probability in the HOM processes. The curves for the pair of detections D$_{1}$D$_{4}$ and D$_{2}$D$_{3}$ coincide in this graph.} 
\end{figure}

\begin{figure}[ht]
\begin{center}
\includegraphics[height=2in]{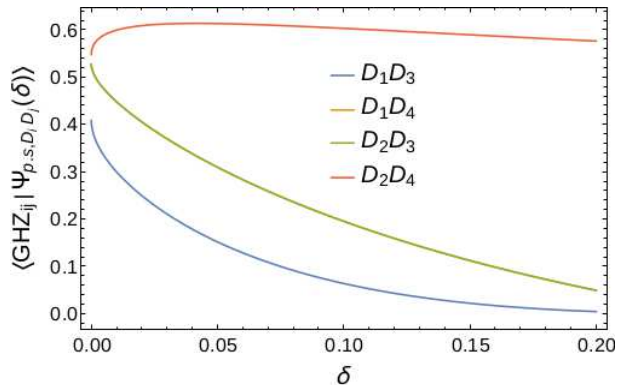}
\end{center}
\label{ogd}
\caption{Overlaps of the different GHZ state outcomes with the corresponding postselected states, i.e., the fidelity in having the desired GHZ states after postselection with the required clicks. The orange and green curves are exactly on top of each othet. One can see that the generation of GHZ states on demand is not possible even if the possible failures in the HOM processes are excluded. The curves for the pair of detections D$_{1}$D$_{4}$ and D$_{2}$D$_{3}$ coincide in this graph.} 
\end{figure}

\section{Conclusions}

In conclusion, we theoretically proposed an optical quantum information protocol intended to entangle three independent, spatially separated photons, using only linear optical elements. To do so, it is required to have five photons, three to become entangled and two to mediate the correlations. The generation of GHZ states is probabilistic, however, it was found that in all successful runs of the protocol genuine tripartite entangled states are created, all pertaining to the GHZ class. Among such states, the original GHZ state can be generated with 5\% of probability, which is considered high if compared to recent proposals \cite{berg}. Possibly, the main building block for the experimental realization of the scheme is the Hong-Ou-Mandel effect which has been observed and manipulated in the laboratory in a number of scenarios \cite{pan2}, including atomic systems \cite{lopes,kauf}. Thus, we believe that our proposal can be experimentally realized with current technology.



\section*{Acknowledgements}

The authors acknowledge the Brazilian funding agency CNPq (AC's Universal grant No. 423713/2016-7, BLB's PQ grant No. 309292/2016-6), 
UFAL (AC's paid license for scientific cooperation at UFRN),
MEC/UFRN (postdoctoral fellowships at IIP).

We also thank Rafael Chaves for fruitful discussions.

\section*{Appendix: Calculating the error influence in the protocol of GHZ state generation}

Here we show how to obtain the most general outcome in our protocol of GHZ state generation, as shown in Fig. 1 of the main article. First, one needs to consider all possible paths taken by the photons. At this point, we shall treat each photon independently. For the sake of notation, we divide the interferometer of Fig. 1 into $10$ regions (channels), which represents the upper and lower halves of the five circuits. The upper halves of the original circuits of photons 1 to 5 will be labeled as 1, 3, 5, 7 and 9, respectively. On the other hand, the lower halves of these circuits will be labeled as 2, 4, 6, 8 and 10, respectively. For example, the history reflection-reflection-transmission (r,r,t) of photon 1 can be denoted as $(i)^2 R_1|1,1,2\rangle_1$. The parameter $R_{1}$ represents the reflection amplitude of BS$_{1}$ and $i^{2}$ is the phase acquired due to the two reflections.

Note that photons 1 and 5 can have four different histories, whereas photons 2, 3 and 4 can take six different paths before detection. Therefore, the general outcome has $4^2 6^3=3456$ terms. For example, there is a term
\begin{equation}
|1,1,2\rangle_1|3,3,3\rangle_2|6,6,5\rangle_3|8,8,7\rangle_4 |10,10,9\rangle_5,
\end{equation}
with coefficient
\begin{equation}
i^2 R_1 i^3 R_2 R_6 R_{10}  i T_3 R_8  i T_4 R_9 T_{11}  T_5 i,
\end{equation}
representing the path (r,r,t); (r,r,r), (t,r,t); (t,r,t) and (t,r,t) of photons 1 to 5, respectively. Again, we put a factor $i$ whenever a reflection takes place, and $R_{j}$ ($T_{j}$) is the reflection (transmission) amplitude of BS$_{j}$. This particular path provides single photons in the channels $2, 3, 5, 7, 9$. We denote such outcome as $(2,3,5,7,9)$, which has the information that photon 1 ends up in channel 2, photon 2 in channel 3, etc. If an experiment had this history, detectors $D_1$ and $D_3$ would click.

Let us consider the following two histories
\begin{equation}
|2,2,1\rangle_1|3,3,3\rangle_2|6,7,8\rangle_3|8,8,10\rangle_4 |10,10,9\rangle_5,
\end{equation}
with coefficient
\begin{equation}
-i R_6^2 R_2 R_{10} T_1 T_3 T_8 T_{11} T_4 T_9 T_5,
\end{equation}
and
\begin{equation}
|2,3,3\rangle_1|3,2,1\rangle_2|6,7,8\rangle_3|8,8,10\rangle_4 |10,10,9\rangle_5,
\end{equation}
with coefficient
\begin{equation}
i T_6^2 R_2 R_{10} T_1 T_3 T_8 T_{11} T_4 T_9 T_5.
\end{equation}
The beam splitters in our proposal have $R=T=1/\sqrt{2}$. If the photons are \textit{indistinguishable}, the above two histories are equivalent. Due to the opposite sign in the linear combination, this term does not have any contribution at the end (the HOM effect at BS$_6$ with the simultaneously arrivals of photons $1$ and $2$). One can handle the failure of the HOM effect at the second layer by keeping the photons \textit{distinguishable}, but changing $R_i^2=T_i^2=\delta$ for $i=6\dots9$ in the coefficients, and then setting the remaining factors $R_i$ and $T_i$ to $1/\sqrt{2}$ (50:50 beam splitters and allowed failure of the HOM effect with probability $\delta$). When $\delta=0$ we have the ideal case considered in the main text.

Having the above substitutions been carried out, one can apply detector projections. Here we suppose that the detectors can count the number of incoming photons. To get the GHZ state ($|\psi\rangle_1$), for example, one needs exactly one click on the detectors $D_1$ and $D_3$, and no click on the other two detectors (postselection). Therefore, from the general linear combination of the outgoing state, one can select only the ones which have exactly one photon in channels $3$ and $7$, and no photons in channels $4$ and $8$.

As a matter of fact, we can restore indistinguishability of the photons by checking the last parts of the history and identifying the indistinguishable outcomes. For example, one identifies $(1,3,8,10,9)$ with $(3,1,8,10,9)$. See the HOM failure example above. Then, one has to sum up over all the possible histories generating the same outcome.
With $\delta=0$, the $D_1D_3$ postselection leads to the following general outcome:
\begin{widetext}

\begin{eqnarray}
|\psi(\delta=0)\rangle_{1,gen}=\frac{1}{64} i |1\rangle _1 |1\rangle _2 |5\rangle _4+\frac{1}{64} i |1\rangle _1 |5\rangle _3 |5\rangle _4+\frac{1}{64} |6\rangle _2 |5\rangle _3 |5\rangle _4 + \frac{i |2\rangle _1 |6\rangle _3 |5\rangle _4}{32 \sqrt{2}}+\frac{1}{64} i |6\rangle _2 |6\rangle _3 |5\rangle _4-\frac{|2\rangle _1 |1\rangle _2 |5\rangle _4}{32 \sqrt{2}}\nonumber \\
+\frac{i |1\rangle _1 |5\rangle _3 |9\rangle _5}{32 \sqrt{2}}+\frac{|6\rangle _2 |5\rangle _3 |9\rangle _5}{32 \sqrt{2}}+\frac{|1\rangle _1 |10\rangle _4 |9\rangle _5}{32 \sqrt{2}}+\frac{1}{64} |1\rangle _1 |1\rangle _2 |10\rangle _5+\frac{i |2\rangle _1 |1\rangle _2 |10\rangle _5}{32 \sqrt{2}}+\frac{|2\rangle _1 |6\rangle _3 |10\rangle _5}{32 \sqrt{2}}\nonumber \\
+\frac{1}{64} |6\rangle _2 |6\rangle _3 |10\rangle _5+\frac{1}{64} i |1\rangle _1 |10\rangle _4 |10\rangle _5+\frac{1}{64} |6\rangle _2 |10\rangle _4 |10\rangle _5-\frac{i |6\rangle _2 |9\rangle _5 |10\rangle _4}{32 \sqrt{2}}. \nonumber \\
\end{eqnarray}

\end{widetext}

If $\delta\neq0$, one has even more terms. One can easily see that there are many outcomes with multiple photons in the same channel.

Let us suppose that one can make an extra post-selection step, namely, selecting outcomes with single photons in each channel (for example, by using at the end an apparatus which works only with single photon input states). In this case, the general ($\delta\neq0$) outcome is the following

\begin{eqnarray}
\frac{i \left(1-2 \sqrt{\delta }\right)^2 |1\rangle _1 |5\rangle _3 |9\rangle _5}{32 \sqrt{2}}+\frac{\left(1-2 \sqrt{\delta }\right)^2 |2\rangle _1 |6\rangle _3 |10\rangle _5}{32 \sqrt{2}}+\frac{1}{32} \sqrt{\delta } \left(1-2 \sqrt{\delta }\right) |1\rangle _1 |5\rangle _3 |10\rangle _4+\frac{\sqrt{\delta } \left(1-2 \sqrt{\delta }\right) |2\rangle _1 |6\rangle _3 |10\rangle _4}{16 \sqrt{2}} \nonumber \\
+\frac{\sqrt{\delta } \left(1-2 \sqrt{\delta }\right) |2\rangle _1 |6\rangle _2 |10\rangle _5}{16 \sqrt{2}}+\frac{i \delta  |1\rangle _2 |5\rangle _4 |9\rangle _5}{8 \sqrt{2}}+\frac{1}{16} \delta  |1\rangle _2 |5\rangle _3 |10\rangle _4+\frac{1}{16} \delta  \left(\sqrt{2} |2\rangle _1-i |1\rangle _1\right) |6\rangle _2 |10\rangle _4 \nonumber \\
+\frac{1}{16} i \delta  |1\rangle _2 |6\rangle _3 |10\rangle _4+\frac{1}{32} \left(2 \sqrt{\delta }-1\right) \sqrt{\delta } |1\rangle _1 |5\rangle _4 |10\rangle _5-\frac{1}{16} \delta  |1\rangle _2 |5\rangle _4 |10\rangle _5+\frac{1}{32} i \left(2 \sqrt{\delta }-1\right) \sqrt{\delta } |1\rangle _1 |6\rangle _2 |10\rangle _5 \nonumber \\
-\frac{1}{32} i \left(2 \sqrt{\delta }-1\right) \sqrt{\delta } |1\rangle _2 |6\rangle _3 |10\rangle _5-\frac{i \left(2 \sqrt{\delta }-1\right) \sqrt{\delta } |1\rangle _2 |5\rangle _3 |9\rangle _5}{16 \sqrt{2}}-\frac{i \left(2 \sqrt{\delta }-1\right) \sqrt{\delta } |1\rangle _1 |5\rangle _4 |9\rangle _5}{16 \sqrt{2}} \nonumber \\
\end{eqnarray}

It is easy to see that the GHZ state is obtained when $\delta=0$ 
\begin{eqnarray}
\frac{|2\rangle _1 |6\rangle _3 |10\rangle _5}{32 \sqrt{2}}+\frac{i |1\rangle _1 |5\rangle _3 |9\rangle _5}{32 \sqrt{2}},
\end{eqnarray}
or restoring the notation of Section II. in the main text
\begin{eqnarray}
\frac{|B\rangle _1 |B\rangle _3 |B\rangle _5}{32 \sqrt{2}}+\frac{i |A\rangle _1 |A\rangle _3 |A\rangle _5}{32 \sqrt{2}}.
\end{eqnarray}

\bibliography{refs2}

\end{document}